\documentclass[twocolumn]{aastex701}

\usepackage{graphicx}
\usepackage{txfonts}
\usepackage{hyperref}
\usepackage{natbib}
\usepackage{float}
\usepackage{layout}
\usepackage{booktabs}
\usepackage{textgreek}
\usepackage{booktabs}

\newcommand{\sophi}{{\sc SO/PHI}}
\newcommand{\hrt}{{\sc SO/PHI-HRT}}
\newcommand{\solo}{Solar Orbiter}

\newcommand{\degree}{$^\circ$}

\begin{document}

\title{The first out-of-ecliptic observations of the polar magnetic field of the Sun}

\author[orcid=0000-0003-2755-5295]{D.~Calchetti}
\affiliation{Max-Planck-Institut f\"ur Sonnensystemforschung, Justus-von-Liebig-Weg 3, 37077 G\"ottingen, Germany}
\email[show]{calchetti@mps.mpg.de}  

\author[0000-0002-3418-8449]{S. K. Solanki}
\affiliation{Max-Planck-Institut f\"ur Sonnensystemforschung, Justus-von-Liebig-Weg 3, 37077 G\"ottingen, Germany}
\email{solanki@mps.mpg.de}

\author{J. Hirzberger}
\affiliation{Max-Planck-Institut f\"ur Sonnensystemforschung, Justus-von-Liebig-Weg 3, 37077 G\"ottingen, Germany}
\email{hirzberger@mps.mpg.de}

\author[0000-0001-7809-0067]{G. Valori}
\affiliation{Max-Planck-Institut f\"ur Sonnensystemforschung, Justus-von-Liebig-Weg 3, 37077 G\"ottingen, Germany}
\email{valori@mps.mpg.de}

\author[0000-0002-9270-6785]{L. P. Chitta}
\affiliation{Max-Planck-Institut f\"ur Sonnensystemforschung, Justus-von-Liebig-Weg 3, 37077 G\"ottingen, Germany}
\email{chitta@mps.mpg.de}  

\author[0000-0002-2055-441X]{J. Blanco Rodr\'\i guez}
\affiliation{Universitat de Val\`encia, Catedr\'atico Jos\'e Beltr\'an 2, E-46980 Paterna-Valencia, Spain}
\affiliation{Spanish Space Solar Physics Consortium (S$^{3}$PC), Spain}
\email{julian.blanco@uv.es}

\author[0000-0002-4693-1156]{A. Giunta}
\affiliation{RAL Space, UKRI STFC Rutherford Appleton Laboratory, Didcot OX11 0QX, UK}
\affiliation{University of Catania, Astrophysics Section, Dept. of Physics and Astronomy, Via S. Sofia 78, 95123 Catania, Italy}
\email{alessandra.giunta@stfc.ac.uk}

\author[]{T. Grundy}
\affiliation{RAL Space, UKRI STFC Rutherford Appleton Laboratory, Didcot OX11 0QX, UK}
\email{tim.grundy@stfc.ac.uk}

\author[0000-0002-3776-9548]{K. Albert}
\affiliation{Max-Planck-Institut f\"ur Sonnensystemforschung, Justus-von-Liebig-Weg 3, 37077 G\"ottingen, Germany}
\email{albert@mps.mpg.de}

\author[0000-0002-1790-1951]{T. Appourchaux}
\affiliation{Univ. Paris-Saclay, Institut d’Astrophysique Spatiale, UMR 8617, CNRS, B{\^a}timent 121, 91405 Orsay Cedex, France}
\email{Thierry.Appourchaux@universite-paris-saclay.fr}

\author[0000-0002-7318-3536]{F. J. Bail{\'e}n}
\affiliation{Instituto de Astrofísica de Andalucía (IAA-CSIC), Apartado de Correos 3004, E-18080 Granada, Spain}
\affiliation{Spanish Space Solar Physics Consortium (S$^{3}$PC), Spain}
\email{fbailen@iaa.es}

\author[0000-0001-8669-8857]{L. R. Bellot~Rubio}
\affiliation{Instituto de Astrofísica de Andalucía (IAA-CSIC), Apartado de Correos 3004, E-18080 Granada, Spain}
\affiliation{Spanish Space Solar Physics Consortium (S$^{3}$PC), Spain}
\email{lbellot@iaa.es}

\author[]{A. Feller}
\affiliation{Max-Planck-Institut f\"ur Sonnensystemforschung, Justus-von-Liebig-Weg 3, 37077 G\"ottingen, Germany}
\email{feller@mps.mpg.de}

\author[0000-0002-9972-9840]{A. Gandorfer}
\affiliation{Max-Planck-Institut f\"ur Sonnensystemforschung, Justus-von-Liebig-Weg 3, 37077 G\"ottingen, Germany}
\email{gandorfer@mps.mpg.de}

\author[0000-0001-7696-8665]{L. Gizon}
\affiliation{Max-Planck-Institut f\"ur Sonnensystemforschung, Justus-von-Liebig-Weg 3, 37077 G\"ottingen, Germany}
\affiliation{Institut f\"ur Astrophysik, Georg-August-Universit\"at G\"ottingen, Friedrich-Hund-Platz 1, 37077 G\"ottingen, Germany}
\email{gizon@mps.mpg.de}

\author[0000-0003-1459-7074]{A. Korpi-Lagg}
\affiliation{Max-Planck-Institut f\"ur Sonnensystemforschung, Justus-von-Liebig-Weg 3, 37077 G\"ottingen, Germany}
\email{lagg@mps.mpg.de}

\author[0000-0001-8164-5633]{X. Li}
\affiliation{Max-Planck-Institut f\"ur Sonnensystemforschung, Justus-von-Liebig-Weg 3, 37077 G\"ottingen, Germany}
\email{lixiaohong@mps.mpg.de}

\author[0000-0002-7336-0926]{A. Moreno Vacas}
\affiliation{Instituto de Astrofísica de Andalucía (IAA-CSIC), Apartado de Correos 3004, E-18080 Granada, Spain}
\affiliation{Spanish Space Solar Physics Consortium (S$^{3}$PC), Spain}
\email{amoreno@iaa.es}

\author[0000-0002-7044-6281]{T. Oba}
\affiliation{Max-Planck-Institut f\"ur Sonnensystemforschung, Justus-von-Liebig-Weg 3, 37077 G\"ottingen, Germany}
\email{oba@mps.mpg.de}

\author[0000-0001-8829-1938]{D. Orozco~Su\'arez}
\affiliation{Instituto de Astrofísica de Andalucía (IAA-CSIC), Apartado de Correos 3004, E-18080 Granada, Spain}
\affiliation{Spanish Space Solar Physics Consortium (S$^{3}$PC), Spain}
\email{orozco@iaa.es}

\author[0000-0002-2391-6156]{J. Schou}
\affiliation{Max-Planck-Institut f\"ur Sonnensystemforschung, Justus-von-Liebig-Weg 3, 37077 G\"ottingen, Germany}
\email{schou@mps.mpg.de}

\author[0000-0001-6060-9078]{U. Sch\" uhle}
\affiliation{Max-Planck-Institut f\"ur Sonnensystemforschung, Justus-von-Liebig-Weg 3, 37077 G\"ottingen, Germany}
\email{schuehle@mps.mpg.de}

\author[0000-0002-5387-636X]{J. Sinjan}
\affiliation{Max-Planck-Institut f\"ur Sonnensystemforschung, Justus-von-Liebig-Weg 3, 37077 G\"ottingen, Germany}
\email{sinjan@mps.mpg.de}

\author[0000-0003-1483-4535]{H. Strecker}
\affiliation{Instituto de Astrofísica de Andalucía (IAA-CSIC), Apartado de Correos 3004, E-18080 Granada, Spain}
\affiliation{Spanish Space Solar Physics Consortium (S$^{3}$PC), Spain}
\email{streckerh@iaa.es}

\author[0000-0002-3387-026X]{J. C. del Toro Iniesta}
\affiliation{Instituto de Astrofísica de Andalucía (IAA-CSIC), Apartado de Correos 3004, E-18080 Granada, Spain}
\affiliation{Spanish Space Solar Physics Consortium (S$^{3}$PC), Spain}
\email{jti@iaa.es}

\author[0000-0002-1217-6455]{A. Ulyanov}
\affiliation{Max-Planck-Institut f\"ur Sonnensystemforschung, Justus-von-Liebig-Weg 3, 37077 G\"ottingen, Germany}
\email{ulyanov@mps.mpg.de}

\author[]{R. Volkmer}
\affiliation{Institut f\"{u}r Sonnenphysik (KIS), Georges-K\"{o}hler-Allee 401a, 79110 Freiburg, Germany}
\email{volkmer@leibniz-kis.de}

\author[0000-0001-5833-3738]{J. Woch}
\affiliation{Max-Planck-Institut f\"ur Sonnensystemforschung, Justus-von-Liebig-Weg 3, 37077 G\"ottingen, Germany}
\email{woch@mps.mpg.de}

\begin{abstract}
Direct remote-sensing observations of the solar poles have been hindered by the restricted view obtained from the ecliptic plane. 
For the first time ever, Solar Orbiter with its remote-sensing instruments observed the poles of the Sun from out of the ecliptic in the Spring of 2025. 
Here we report the first measurements of the magnetic field of the solar poles taken when Solar Orbiter was at heliographic latitudes ranging between 14.9\degree and 16.7\degree. 
The data-sets were collected by the High Resolution Telescope of the Polarimetric and Helioseismic Imager (SO/PHI-HRT) on board Solar Orbiter. 
Two sets of observations, approximately one month apart, for the south and north pole are considered in this work. 
The magnetic flux and flux density measured during these campaigns are reported as a function of the heliographic latitude observed by \hrt. 
The net fluxes show a different latitudinal distribution for the two polar caps. 
We also discuss the observed dependence of the measured fluxes on the viewing angle. 
These first results highlight the importance of high-resolution direct measurements of the polar field, paving the way to the high-latitude observations planned for SO/PHI-HRT in the coming years. 
\end{abstract}


\section{Introduction}\label{sec:intro}
\begin{figure*}[t]
    \centering
    \includegraphics[width=0.98\textwidth]{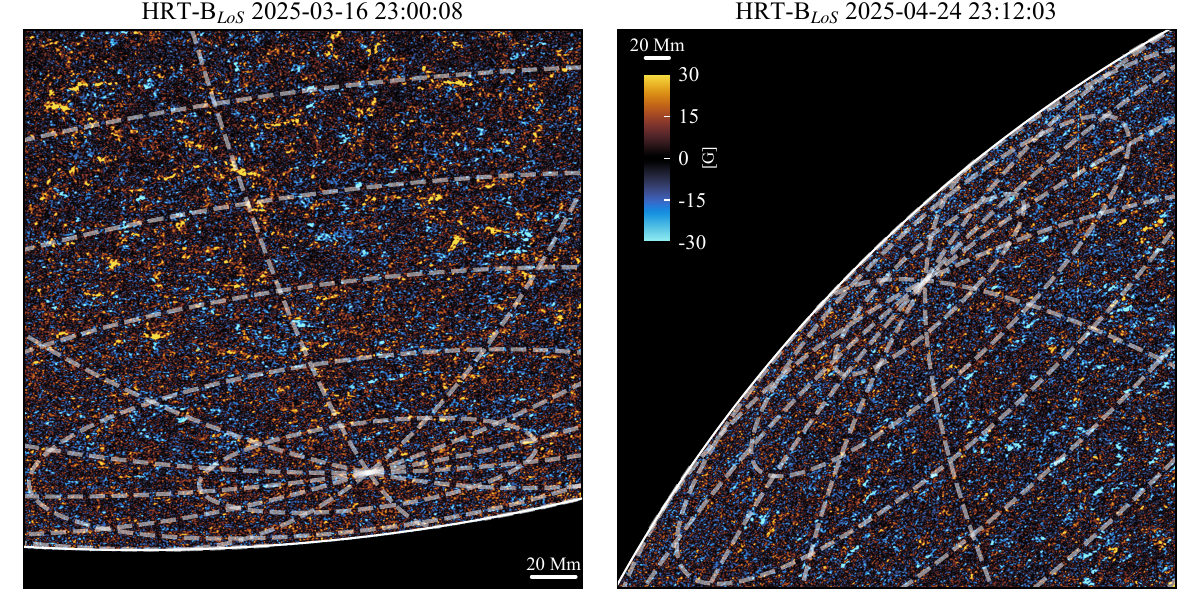}
    \caption{Magnetic field at the solar poles. Partial field-of-view of the line-of-sight magnetic field acquired during the south (left panel) and north (right panel) polar campaigns. The dashed lines show the heliographic longitudes and latitudes with 30\degree\ and 5\degree\ spacing respectively. The acquisition time is specified on top of each panel. The orientation of the field-of-view is caused by the spacecraft roll. An animation of this figure is available online. The animation has a playback time of 16~s and shows the full field-of-view of the data acquired during both campaigns. }
    \label{fig:data}
\end{figure*}
The Sun exhibits a magnetic activity cycle of approximately 11 years. 
During the solar minimum, the magnetic field of the south and north poles shows opposite polarities. 
The initially unipolar caps evolve into mixed-polarity regions as the activity increases, until a magnetic field reversal occurs between the maximum of activity and the next solar minimum \citep{1961ApJ...133..572B,1969ApJ...156....1L,1989ApJ...347..529W,2017SSRv..210...77P}. 
The unipolar caps play a key role in determining the flux of the Sun's open magnetic field \citep{2017ApJ...848...70L}, the interplanetary magnetic field at 1~au \citep[IMF,][]{2012LRSP....9....6M}, and the origin of the fast solar wind \citep{1973SoPh...29..505K,2005Sci...308..519T,2023Sci...381..867C}. 
The magnetic field at the poles is also thought to provide the seed for the next solar cycle \citep{2020LRSP...17....4C,2016ApJ...823L..22C}. 

The first high-resolution observations of the magnetic field vector were carried out from ground-based observatories by \cite{1997SoPh..175...81H}, \cite{2004A&A...425..321O}, and \cite{2007A&A...474..251B}. 
\cite{2008ApJ...688.1374T} for the first time revealed the polar magnetic landscape observed at high resolution from space, with the Spectro-Polarimeter \citep[SP,][]{2013SoPh..283..579L} of the Solar Optical Telescope \citep[SOT,][]{2008SoPh..249..167T} onboard Hinode \citep{2007SoPh..243....3K}. 
The data acquired by Hinode have been extensively used to track the evolution of the magnetic field at the poles, including its reversal \citep{shiota2012,2024RAA....24g5015Y,2022ApJ...941..142P}. 

Observations of the solar magnetic field vector at the polar regions has proven to be extremely challenging for three reasons. 
Firstly, foreshortening of the field-of-view affects the spatial resolution of the measurement. 
This in turn leads to increased cancellation of Stokes $V$ signals of opposite polarity magnetic fields \citep[e.g.][]{2004A&A...417.1125K} and reduces the detectability of the smaller polar faculae. 
Secondly, the lower intensity measured at the limb, due to the limb darkening \citep{1977SoPh...51...25P}, reduces drastically the signal-to-noise ratio (SNR) of the data. 
Thirdly, the geometry of the observation complicates the interpretation of the results compared to disk-center observations, primarily because very close to the limb, the radial component of the magnetic field is best inferred from linear polarization signals, which are generally weaker and more susceptible to noise than circular polarization \citep{2017SoPh..292...13P,2022ApJ...941..142P}.  
All these effects make the polar magnetic field, particularly its line-of-sight component, difficult to measure and to interpret \citep{centeno2023}. 

One way to overcome these challenges is to obtain observations of the magnetic field from higher heliographic latitudes than those accessible from the near-Earth environment observatories. 
This has now become possible, for the first time, with the Polarimetric and Helioseismic Imager on board Solar Orbiter \citep[\sophi][]{solo,2020A&A...642A..11S}. 
In February 2025, Solar Orbiter entered an orbit which allows its remote sensing instruments to image the Sun from up to $\pm$17\degree in heliographic latitude, 10\degree\ more than what is achievable from Earth. 

In this paper, we report the first remote-sensing observation of the solar north and south pole magnetic field from out of the ecliptic plane taken with the \sophi\ High Resolution Telescope \citep[\hrt][]{gandorfer2018high}. 
The data used in this work are described in Section~\ref{sec:data}. 
The results obtained during the observing campaigns are reported in Section~\ref{sec:result}. 
The discussion of the results and a brief summary are given in Section~\ref{sec:end}. 
%
%
\section{Data}\label{sec:data}

Data from two campaigns are analyzed to investigate both the south and north poles of the Sun. 
The south pole observations were taken during the Solar Orbiter Observing Plan (SOOP) R\_SMALL\_HRES\_MCAD\_Polar-Rotation \citep{2020A&A...642A...3Z} between 16 and 17 March 2025. 
The time series spans 42 hours of observation at 1-hour cadence, pointing close to the solar south pole when Solar Orbiter was approaching the Sun moving from -14.9\degree\ to -15.6\degree\ in heliographic latitude and from a distance of 0.43 to 0.41~au. 
For comparison, the maximum inclination that can be obtained from near-Earth environment observatories is $\pm$7.25\degree. 
During this campaign, two consecutive datasets (100~s apart) are averaged to increase the SNR. 
The north pole observations were acquired during the R\_SMALL\_HRES\_MCAD\_Polar-Observations SOOP between 24 and 25 April 2025. 
The observations consist of eight datasets taken at three different pointings close to the north pole of the Sun. 
During this campaign, Solar Orbiter latitude was at approximately 16.7\degree heliolatitude, at a distance of 0.55~au from the Sun. 
The data are processed with the \hrt\ reduction pipeline \citep{hrt-pipeline,fatima-PD,2024-fran} and the magnetic field vector is retrieved by solving the radiative transfer equation (RTE) with the MILOS code \citep{2007A&A...462.1137O}. 
Maps of the line-of-sight magnetic field for both campaigns, saturated at $\pm$30~G, are shown in Figure~\ref{fig:data}. 
The movie of the campaigns is provided as supplementary material (S1.mp4). 
The fully calibrated Level~2 data used for this work are publicly available in the Solar Orbiter Archive (SOAR). 

To avoid spurious magnetic signal coming from noisy regions, the magnetic elements are masked using the circular and linear polarization signals following the approach of \cite{2020A&A...644A..86P}. 
Specifically, we select all points with circular polarization $|CP - \alpha_{CP}| > 2\sigma_{CP}$ 
or linear polarization $LP - \alpha_{LP} > 2\sigma_{LP}$, 
where $\alpha_{CP,LP}$ is the average and $\sigma_{CP,LP}$ is the standard deviation of the circular and linear polarization, respectively, determined from a Gaussian fit of the distribution of the polarization values. 
Then, all elements with a size of at least 4~pixels and at least one pixel with circular or linear polarization above $3\sigma$ are selected. 

The ambiguity in the magnetic field is resolved using the method developed by \cite{ito2010}. 
The resulting magnetic field vectors are then classified into vertical (positive or negative) fields, horizontal fields, and undetermined fields. 
The method is re-adapted for observations taken from outside the ecliptic, which means that the latitude used in Equations (1) and (2) in \cite{ito2010} corresponds to the latitude as seen from Solar Orbiter rather than the true heliographic latitude. 
%
%
\section{Results}\label{sec:result}
\begin{figure}
    \centering
    \includegraphics[width=0.999\columnwidth]{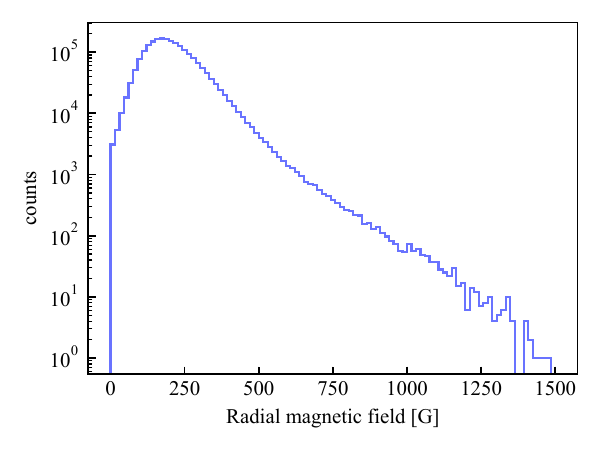}
    \caption{Histogram showing the absolute value of the radial magnetic field measured during the polar campaigns. }
    \label{fig:field-hist}
\end{figure}

In this study we focus on the properties of the radial magnetic field only, as determined by inverting the full magnetic vector.
First, we consider only pixels that satisfy our noise criteria, as defined in Section \ref{sec:data}. 
Second, of the subset of pixels with signal above the noise level, we select only those whose disambiguation provided a predominantly vertical field, i.e. with an angle to the radial direction below 40\degree\ or above 140\degree. 
The values of the radial magnetic field measured inside all the elements selected in both campaigns are shown in Figure~\ref{fig:field-hist}. 
The distribution peaks at $\sim$200~G and shows values up to 1.5~kG. 

\begin{figure}
    \centering
    \includegraphics[width=0.999\columnwidth]{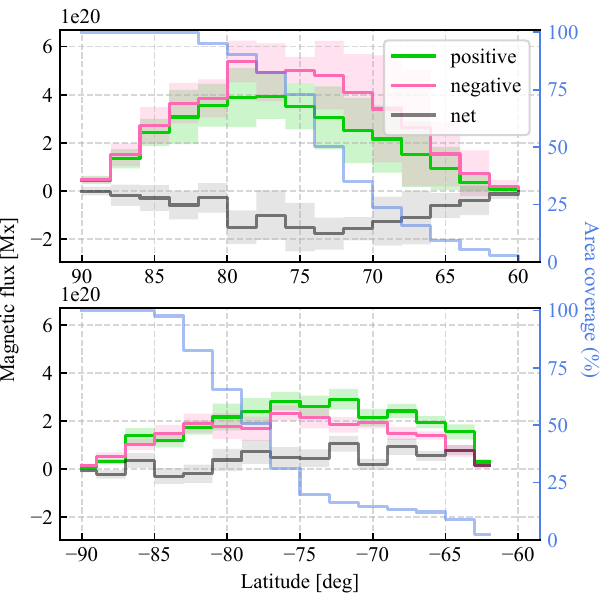}
    \caption{Magnetic fluxes for the north (upper panel) and south pole (bottom panel) campaigns. The green and pink lines show the absolute positive and negative fluxes respectively, and the dark gray line the net flux as a function of the heliographic latitude. The latitudinal bins have a size of 2\degree. The thick lines show the flux averaged over the different observations, the shaded areas show the $1\sigma$ variance between the individual snapshots. The blue line shows the average percentage of area coverage through the campaign at different latitudes. In both panels the respective pole is located on the left side. }
    \label{fig:flux}
\end{figure}

%
The average magnetic flux is calculated for both south and north pole observations and is shown in Figure~\ref{fig:flux} as a function of the heliographic latitude. 
The flux is computed in each pixel of the selected magnetic elements as $B_r\cdot S$, where $B_r$ is the radial component of the magnetic field, obtained after the disambiguation, and $S$ is the surface area of each pixel. 
The magnetic flux is then segmented into latitudinal bins of 2\degree\ size. 
The thick lines represent the fluxes measured within a given latitude bin averaged over longitude and over all data sets of the respective campaigns (42 for the south pole, 8 for the north pole), while the $1\sigma$ variation is shown as a shaded area. 
Both poles show a bell-shaped distribution peaked between 75\degree\ and 70\degree\ with low fluxes both close to the pole and at low absolute latitudes. 
The total flux measured during the campaigns depends also on the area covered by the observations. 
The percentage of the area observed in a specific latitude bin is shown as a blue curve in Figure~\ref{fig:flux}. 
The absolute area of each latitudinal bin is different, which is also affecting the flux measured in that specific bin. 
Indeed, it is more likely to measure higher flux in a large area than in a smaller one. 

\begin{figure}[b]
    \centering
    \includegraphics[width=0.999\columnwidth]{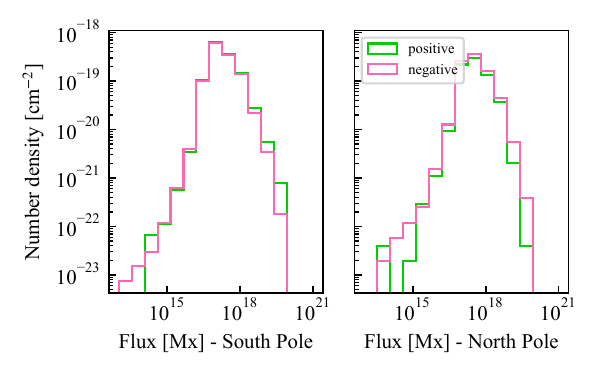}
    \caption{Number density of the magnetic elements as a function of the magnetic fluxes of these elements measured during the south (left panel) and north (right panel) pole campaigns. Only values with absolute heliographic latitude above 70\degree\ are considered.}
    \label{fig:flux-hist}
\end{figure}

The number density of the magnetic elements is displayed in Figure~\ref{fig:flux-hist} as a function of the magnetic flux in individual elements. 
The number of elements in each bin is displayed as a function of the total unsigned flux (similarly to \citealt{shiota2012}). 
The imbalance between different polarities in different ranges of flux is clearly shown in this figure, particularly the north pole shows  negative polarity concentrations with larger flux, whereas the south pole is more balanced, although with some excess of positive polarity flux. 
The differences in the total flux above 70\degree\ in absolute latitude are shown in Table~\ref{tab:flux}. 

\begin{figure}[t]
    \centering
    \includegraphics[width=0.999\columnwidth]{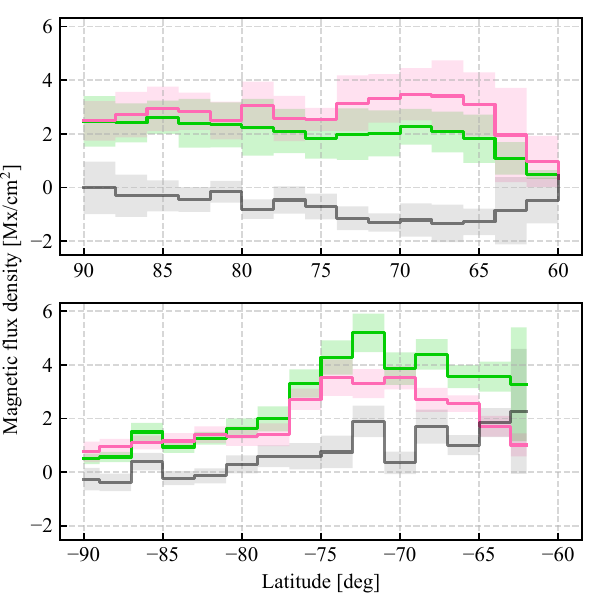}
    \caption{Magnetic flux densities for the north (upper panel) and south pole (bottom panel). The colors are represented as in Figure~\ref{fig:flux}.}
    \label{fig:flux-density}
\end{figure}
\begin{table}[b]
    \centering
    \begin{tabular}{lcc}
        \toprule
        Flux & South Pole & North Pole \\
        \midrule
        Positive & (1.76$\pm$0.29)$\cdot$10$^{21}$ Mx & (2.78$\pm$0.82)$\cdot$10$^{21}$ Mx \\
        \midrule
        Negative & (-1.49$\pm$0.25)$\cdot$10$^{21}$ Mx & (-3.64$\pm$0.75)$\cdot$10$^{21}$ Mx \\
        \midrule
        Net      & (2.75$\pm$ 2.19)$\cdot$10$^{20}$ Mx & (-8.68$\pm$4.13)$\cdot$10$^{20}$ Mx \\
        \bottomrule
    \end{tabular}
    \caption{Averaged fluxes measured during the south and north pole campaigns. The uncertainty corresponds to one standard deviation. Only values with absolute heliographic latitude above 70\degree\ are considered. }
    \label{tab:flux}
\end{table}
Because of the different absolute areas covered by the observations at different latitudes, the magnetic flux density is determined to better evaluate the magnetic fluxes at the poles. 
The flux density is defined as
$$
    B(\lambda) = \frac{\sum_n \Phi(n)}{\sum_{i,j} S_{i,j}}\,,
$$
where $B(\lambda)$ is the magnetic flux density in a specific latitudinal bin $\lambda$, $\Phi(n)$ is the magnetic flux of the $n$-th element in that bin, and $S_{i,j}$ is the surface of each pixel ($i,j$) between the latitudes defined by $\lambda$. 
As for the magnetic flux, the latitudinal bins have a size of 2\degree. 
The result is displayed in Figure~\ref{fig:flux-density}. 
The color scheme is similar to that in Figure~\ref{fig:flux}, but now for the magnetic flux densities, with pink for negative, green for positive and  black for the net magnetic flux density. 

The flux density of both positive and negative features remains roughly constant between 65 and 90\degree\ northern latitudes, while the net flux shows a gradual increase from nearly zero at the pole until a latitude of 65\degree. 
At lower latitudes, the flux density decreases until the edge of the FoV at 60\degree. 
Given the sizes of the error bands, however, all flux densities variations with latitude measured at the north pole are compatible with no dependence on latitude, except for the last bin, 60--65\degree. 
For the south pole, the flux density, particularly the positive one, increases strongly away from the pole until around a  latitude of -70\degree\ after which it stays roughly stable. 
The net magnetic flux density also increases when moving away from the pole, but less pronounced than around the north pole. 
It is noteworthy that the next flux has opposite signs at the two poles.

This result is also illustrated in Table~\ref{tab:flux-density}, where it is shown that the total net magnetic flux density in the north pole is slightly higher than that in the south pole. 
The uncertainties of the north pole are higher with respect to the south pole because of the lower number of datasets acquired in the two campaigns. 

\begin{table}[t]
    \centering
    \begin{tabular}{lcc}
        \toprule
        Flux density & South Pole & North Pole \\
        \midrule
        Positive & (1.99$\pm$0.33) Mx/cm$^2$ & (2.16$\pm$0.80) Mx/cm$^2$ \\
        \midrule
        Negative & (-1.68$\pm$0.29) Mx/cm$^2$ & (-2.82$\pm$0.75) Mx/cm$^2$ \\
        \midrule
        Net      & (0.31 $\pm$ 0.25) Mx/cm$^2$ & (-0.66$\pm$0.32) Mx/cm$^2$ \\
        \bottomrule
    \end{tabular}
    \caption{Temporal average of the flux densities measured during the south and north pole campaigns. The uncertainty corresponds to one standard deviation. Only elements located at absolute heliographic latitudes above 70\degree\ are considered. }
    \label{tab:flux-density}
\end{table}
%
\section{Summary and Discussion}\label{sec:end}
In this work, we reported the first measurement of the magnetic field of the solar poles acquired out of the ecliptic plane by \hrt. 
\solo\ observed both poles approximately one month apart thanks to its highly eccentric and somewhat inclined orbit. 
The main result of the analysis presented here is that the poles already exhibit a polarity reversal compared to the first half of the current solar cycle \citep[see][]{shiota2012,2024RAA....24g5015Y}, as reported in Tables~\ref{tab:flux}~and~\ref{tab:flux-density}. 
Figures~\ref{fig:flux}~and~\ref{fig:flux-density} show the distribution of the magnetic flux and flux density at different latitudes, respectively. 

The region surrounding the south pole displays an excess of positive flux starting from -80\degree\ in heliographic latitude, although in the region immediately surrounding the pole the flux difference is close to the uncertainty in the measurements.  
In contrast, the north pole shows a clear net negative flux between about 65\degree\ in latitude and 85\degree. 
Beyond that latitude the uncertainties are too large to determine if there is a flux imbalance or not. 
In general, the area surrounding the north pole shows higher magnetic flux compared to the south pole and a correspondingly larger imbalance. 
A difference is also seen in the absolute flux density around the two poles, with the flux density at the north pole staying almost constant at a value somewhat in excess of 2~G right up to the pole, while it drops to below 1~G at the south pole. 

The decrease of the magnetic flux density towards the pole is a sign of the delay of polarity reversal at the highest latitudes compared to the lower ones. 
Such an effect reflects the migration of the magnetic flux from lower latitudes, as shown also by \cite{2024RAA....24g5015Y} and \cite{2022ApJ...941..142P}, and modeled by, e.g., \cite{2001ApJ...551.1099S} or \cite{2025A&A...700A.210J}. 
Uncovering the dynamo processes, which in addition leads to the polarity reversal, is one of the main objectives of \solo, of which this work represents a first step to be extended by future, dedicated \sophi\ observation campaigns. 

The uncertainties in the north pole are higher than those in the south pole (see Figure~\ref{fig:flux-density}) because of the lower number of datasets used in the analysis (42 for the south pole, 8 for the north pole) and the lower spatial resolution and SNR. 
As discussed in \cite{2019LRSP...16....1B} and already shown in \cite{2004A&A...417.1125K}, higher resolution helps to reveal previously unresolved magnetic structures, leading to an increase in the measured flux. 
The number density of the magnetic flux in the selected elements (Figure~\ref{fig:flux-hist}) also shows higher values in the south pole than in the north pole for magnetic fluxes below $10^{17}$~Mx.
Such an effect is probably the result of a combination of the lower spatial resolution and SNR in the north pole data which prevents the detection of the smallest (and weakest) elements. 
The lower SNR is due to the type of observations planned for the two campaigns, which allowed the average of two consecutive datasets for the south pole, but not for the north pole. 
The small scale elements are anyway expected to have a limited contribution on the total flux imbalance of the polar caps, as reported by \cite{shiota2012} and \cite{2025arXiv250919739Y}. 

\begin{figure}[!t]
    \centering
    \includegraphics[width=0.99\columnwidth]{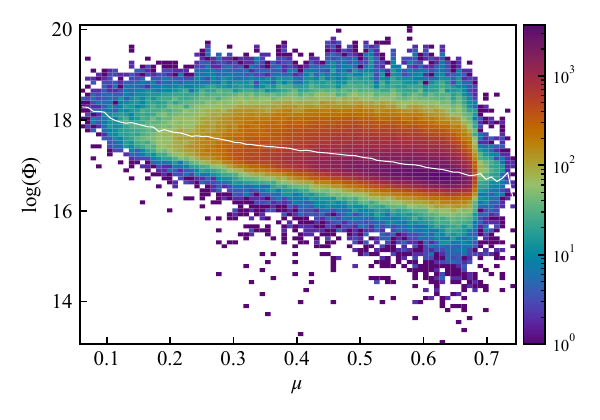}
    \caption{Distribution of the logarithm of the magnetic flux ($y$ axis) of each selected magnetic element at the south and north poles as function of $\mu$ 
    ($x$ axis). The colorbar shows the number of counts in each bin on a logarithmic scale. The white line represents the average of the flux for each $\mu$ bin. }
    \label{fig:flux-mu-lat}
\end{figure}
In addition to the spatial resolution and to the SNR, a crucial factor in the observation of small and weak magnetic patches is the foreshortening close to the solar limb \citep[see also][]{2024A&A...690A.341S, 2025arXiv251000320A}. 
Figure~\ref{fig:flux-mu-lat} shows the dependence of the fluxes of every selected magnetic element on the $\mu = \cos\theta$, where $\theta$ is the heliocentric angle at which they are observed for both the north and south poles. 
The figure clearly shows that the average magnetic flux per feature increases when observing elements close to the limb, because the weaker (and smaller) elements are either not visible or their signal is too small compared to the background noise. 
Furthermore, the elements with the highest fluxes (above $10^{19.5}$ Mx) are observed mostly for $\mu$ values higher than $\sim$0.25. 
At this moment, it is not possible to discern the cause of this distribution, being a foreshortening effect or a real absence of smaller elements at the pole. 
Limb observations far from the poles and future polar campaigns run at even higher latitudes will help point to a solution. 

High latitude observations provide key information to study the solar poles. 
\hrt\ is the only instrument able, now and in the coming five years, to provide magnetic field measurements from out of the ecliptic. 
The ability to image the Sun's high latitude areas at larger $\mu$ will be of fundamental importance for the understanding of the magnetic field at the pole, its generation, its evolution, \and how that shapes the near-Earth environment and the entire heliosphere \citep[see also the study by][that analyzed the poleward migration of the magnetic field at high polar latitudes]{2025ApJ...993L..45C}. 
Our preliminary results set the grounds for such novel science from the high-latitude phase of the Solar Orbiter mission.

%
%

\begin{acknowledgements}
Solar Orbiter is a space mission of international collaboration between ESA and NASA, operated by ESA. We are grateful to the ESA SOC and MOC teams for their support. The German contribution to SO/PHI is funded by the BMWi through DLR and by MPG central funds. The Spanish contribution is funded by AEI/MCIN/10.13039/501100011033/ and European Union ``NextGenerationEU/PRTR'' (RTI2018-096886-C5,  PID2021-125325OB-C5, PID2024-156066OB-C5) and ERDF ``A way of making Europe''; ``Center of Excellence Severo Ochoa'' awards to IAA-CSIC (SEV-2017-0709, CEX2021-001131-S). The French contribution is funded by CNES. 
This project has received funding from the European Research Council (ERC) under the European Union's Horizon 2020 research and innovation programme (grant agreement Nos. 101097844 — project WINSUN; 10103984 -- project ORIGIN). 
This research used version 6.0.6 (\href{https://doi.org/10.5281/zenodo.15690707}{10.5281/zenodo.15690707}) of the SunPy open source software package \citep{sunpy_community2020}.
\end{acknowledgements}
%
%
\bibliographystyle{aasjournalv7}
\bibliography{bibfile}
%
%
\end{document}